\documentclass[usenatbib,a4paper,times,fleqn]{mnras}
\usepackage{graphicx,natbib,times,amssymb,amsmath,pstricks}
\usepackage{aas_macros}
\usepackage{bm,url}
\usepackage{textcomp}
\usepackage{breakurl}
\usepackage{subfig}

\title[Gap opening in dusty discs]{Two mechanisms for dust gap opening in protoplanetary discs}

\author[Dipierro et al.]{Giovanni Dipierro$^{1}$\thanks{giovanni.dipierro@unimi.it}, Guillaume Laibe$^{2}$, Daniel J. Price$^{3}$ and Giuseppe Lodato$^{1}$ \\
$^{1}$Dipartimento di Fisica, Universit\`a Degli Studi di Milano, Via Celoria, 16, Milano, I-20133, Italy \\
$^{2}$School of Physics and Astronomy, University of St. Andrews, North Haugh, St. Andrews, Fife KY16 9SS, UK \\
$^{3}$Monash Centre for Astrophysics (MoCA) and School of Physics and Astronomy, Monash University, Clayton Vic 3800, Australia
}
\pagerange{\pageref{firstpage}--\pageref{lastpage}} \pubyear{2016}
\defcitealias{nsh86}{NSH86}

\date{}
\begin{document}
\label{firstpage}
\bibliographystyle{mnras}
\maketitle

\begin{abstract}
We identify two distinct physical mechanisms for dust gap opening by embedded planets in protoplanetary discs based on the symmetry of the drag-induced motion around the planet: I) A mechanism where low mass planets, that do not disturb the gas, open gaps in dust by tidal torques assisted by drag in the inner disc, but resisted by drag in the outer disc; and II) The usual, drag assisted, mechanism where higher mass planets create pressure maxima in the gas disc which the drag torque then acts to evacuate further in the dust. The first mechanism produces gaps in dust but not gas, while the second produces partial or total gas gaps which are deeper in the dust phase. Dust gaps do not necessarily indicate gas gaps.
  

 \end{abstract}


\section{Introduction}
Recent spectacular spatially resolved observations of gaps and ring-like structures in nearby dusty protoplanetary discs \citep{alma-partnership15a,nomura15a} have revived interest in studying gap-opening mechanisms in discs and in extending the theoretical and numerical investigations conducted over the last four decades \citep[e.g.][]{goldreich79a,goldreich80a,lin86a,lin86b,paardekooper04a,paardekooper06a}. The fundamental question is whether or not these structures are created by embedded protoplanets \citep[e.g.][]{flock15a,zhang15a,dipierro15a,dong15b,jin16a}. 

 However, most of the theory on gaps in discs has considered purely gaseous discs, with dust responding to the gas distribution mainly via drag. As a result, there is one physical mechanism discussed in the literature for explaining the origin of planet-induced gaps and ring-like structures in dusty discs: High mass planets exert a tidal torque that overpowers the local viscous torque and forms a gap in the gas \citep{papaloizou84a,bryden99a}. Large grains then drift towards the pressure maxima located at the gap edges as a result of gas drag. Trapping of dust at the pressure maxima prevents the large grains from being accreted onto the planet and produces a deep structure in the dust  \citep{paardekooper04a,paardekooper06a,fouchet07a,fouchet10a}, easily detectable by ALMA \citep{pinilla12a,pinilla15a,gonzalez12a,gonzalez15a}. Small grains, however, remain strongly coupled to the viscous flow of the gas and are accreted onto the planet \citep{rice06a}.

 Here, motivated by our recent numerical two fluid dust/gas simulations modelling the HL Tau disc \citep{dipierro15a}, we point out that the standard model only partially captures the physics of gap opening in dusty discs. We use our 3D dust-and-gas Smoothed Particle Hydrodynamics (SPH) code to demonstrate the different roles of the tidal and drag torques in opening dust gaps in protoplanetary discs. In particular, we identify regimes for which \emph{it is possible for gaps to form only in the dust, but not in the gas}.
 
 
 

%
\begin{figure*}
\begin{center}
\subfloat{\includegraphics[width=0.32\textwidth]{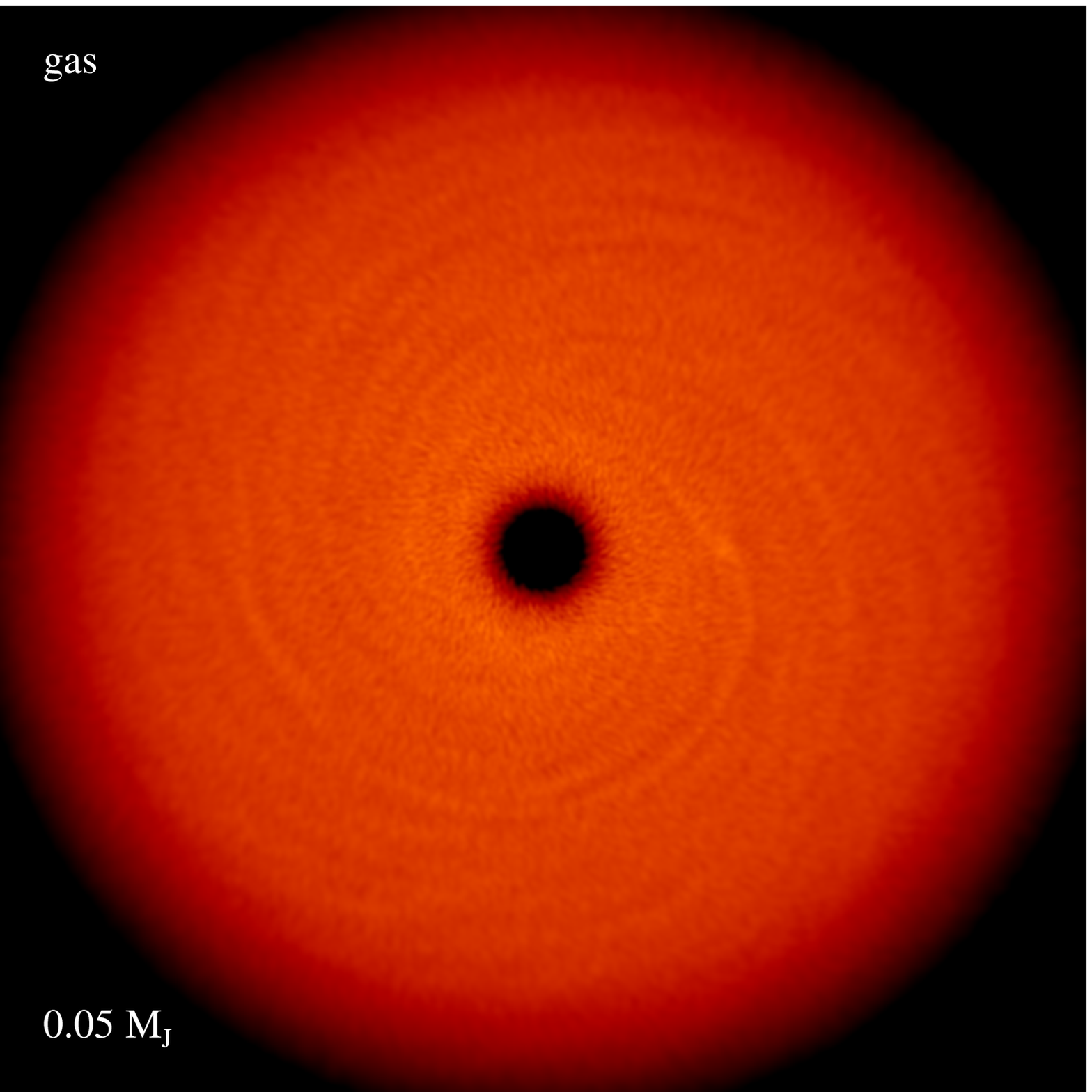}}
\hspace{0.025cm}
\subfloat{\includegraphics[width=0.32\textwidth]{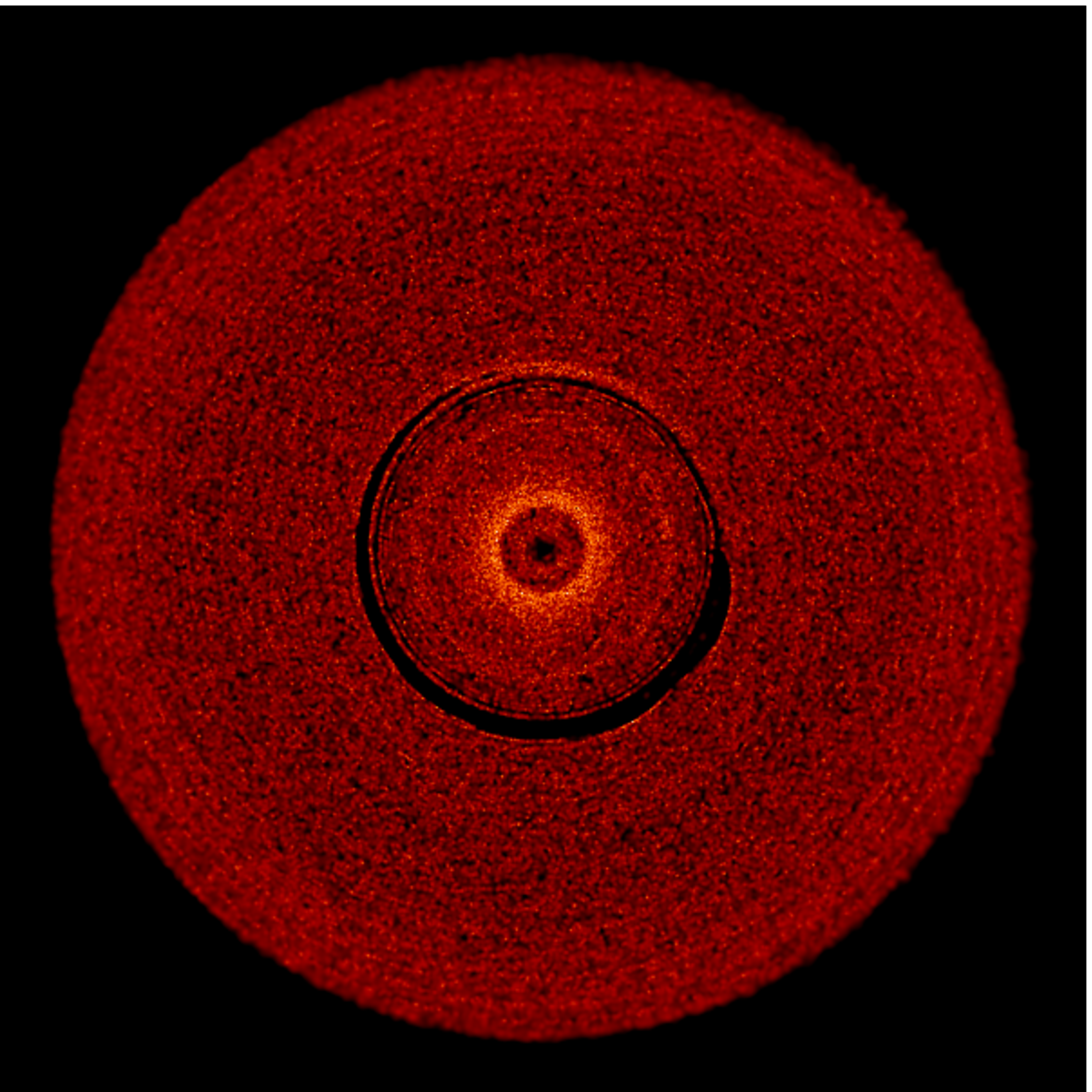}}
\hspace{0.25cm}
\subfloat{\includegraphics[width=0.32\textwidth]{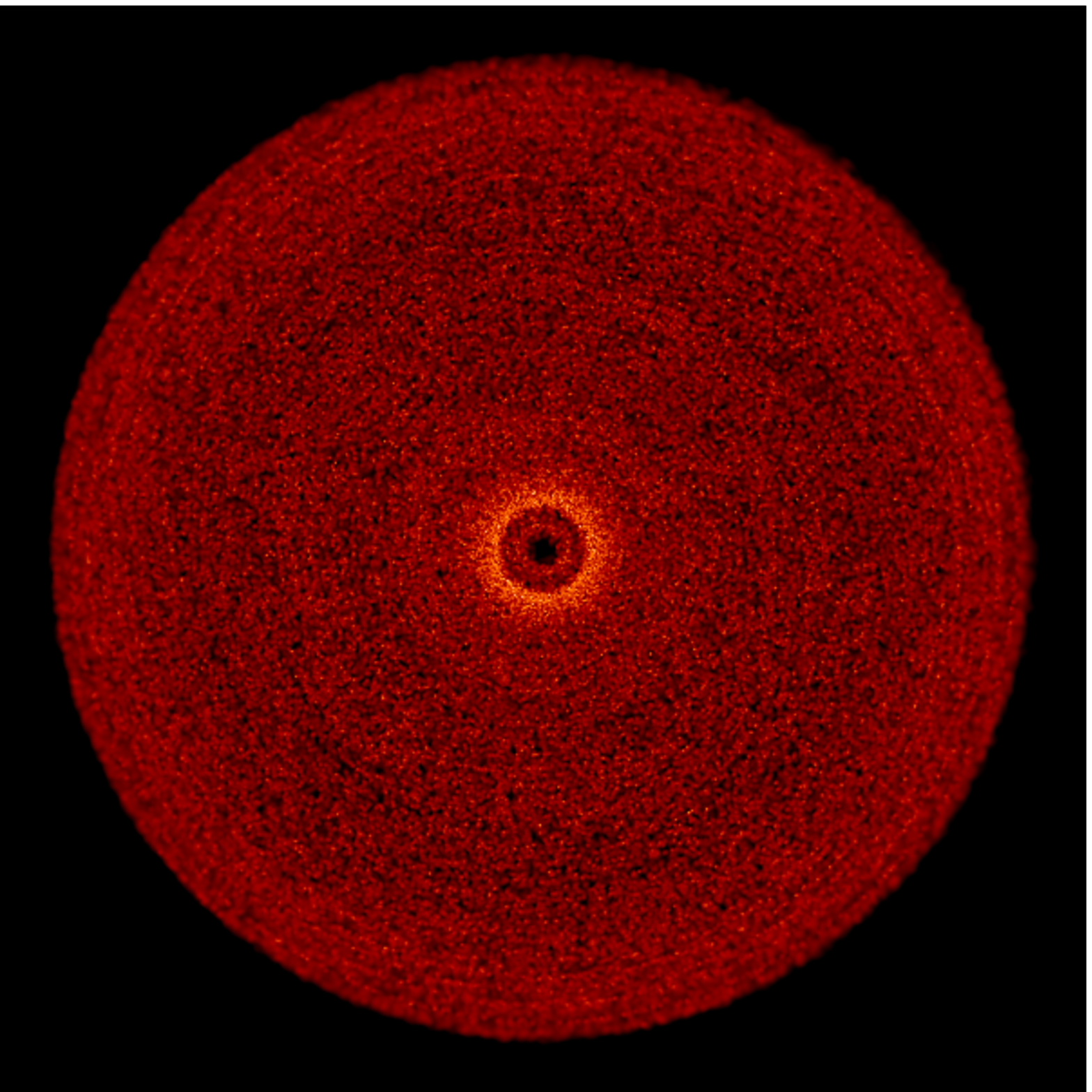}}
\vspace{-0.35cm}
\subfloat{\includegraphics[width=0.32\textwidth]{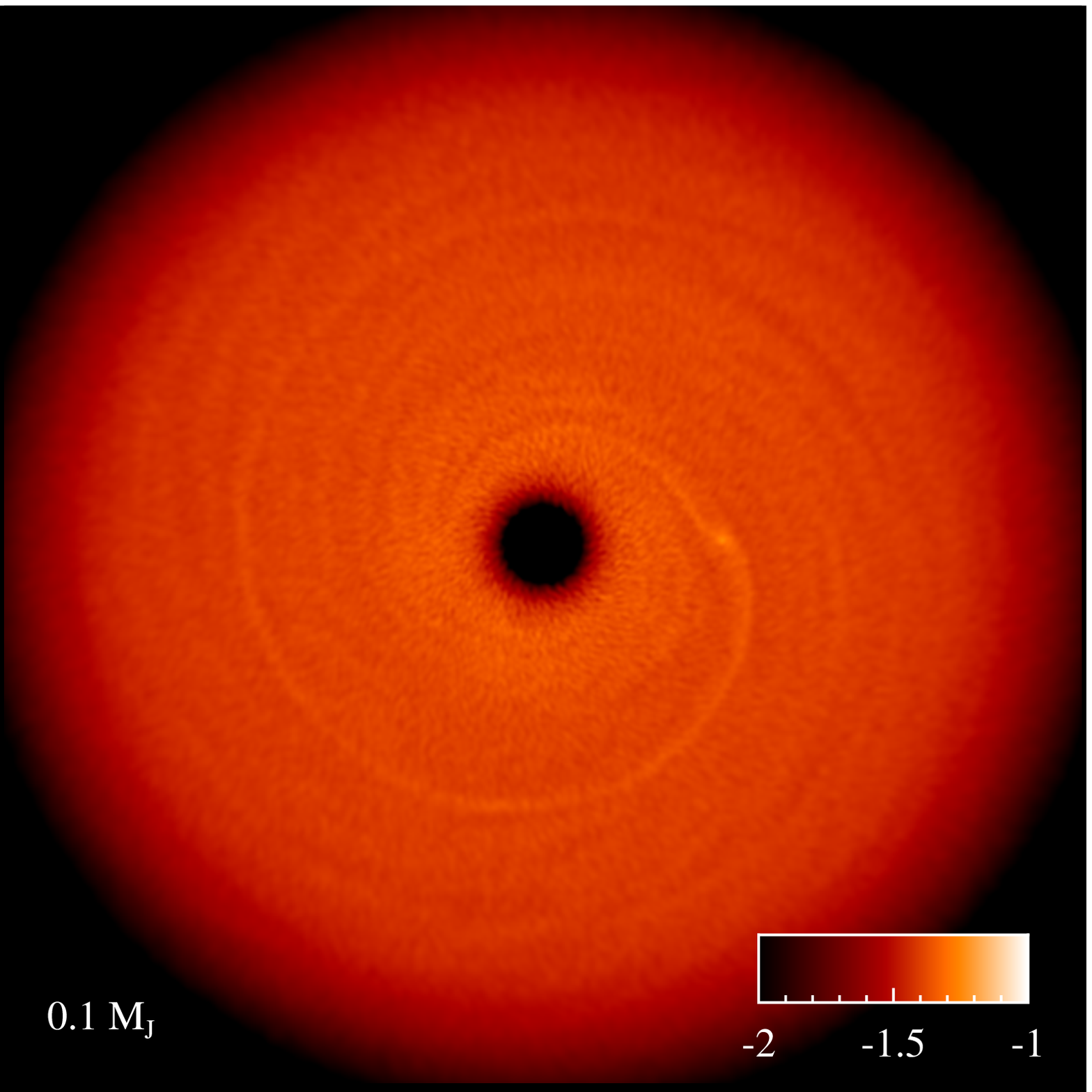}}
\hspace{0.025cm}
\subfloat{\includegraphics[width=0.32\textwidth]{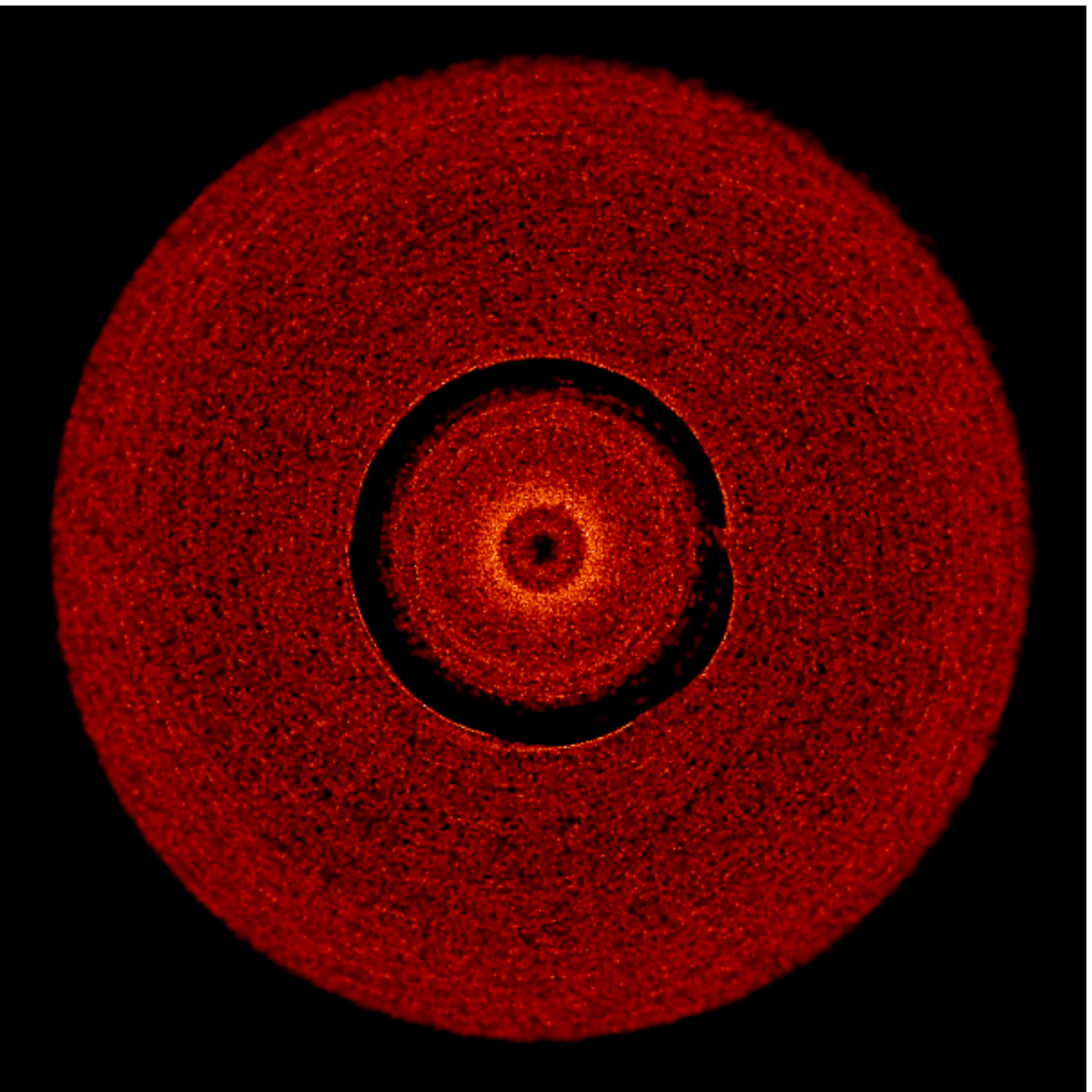}}
\hspace{0.25cm}
\subfloat{\includegraphics[width=0.32\textwidth]{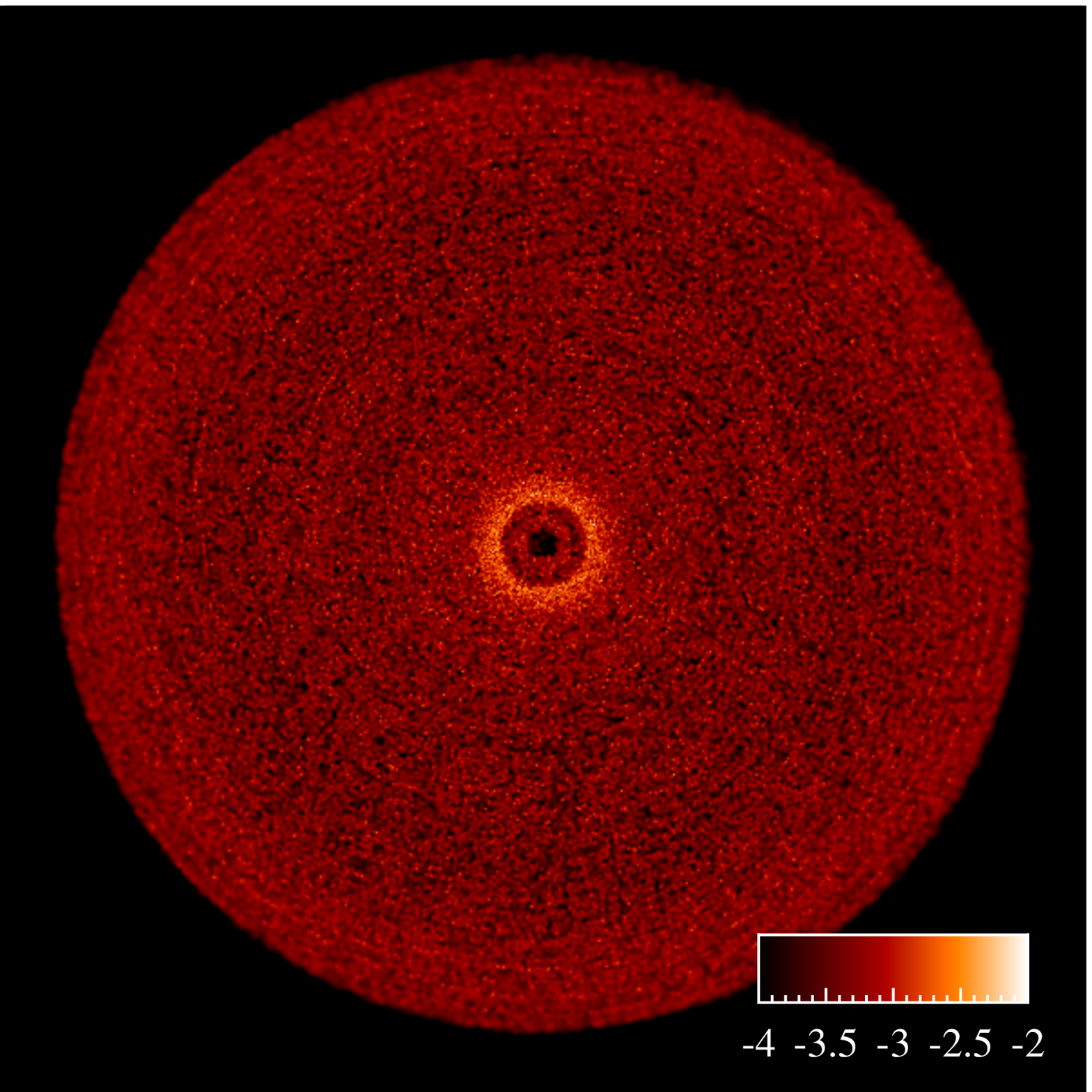}}
\caption{Gap opening via Mechanism I, where low mass planets carve a gap in the dust but not the gas. Plots show gas (left) and millimetre dust grain (centre) surface densities in a dusty disc hosting planets of mass 0.05 $M_{\rm J}$ (top row) and 0.1 $M_{\rm J}$ (bottom row). While the 0.05 $M_{\rm J}$ creates a depletion of dust at the planet location, the 0.1 $M_{\rm J}$ planet is able to carve a gap in the dust. Neglecting the gravity between the planet and the dust (right panels) shows that the gap is opened by the tidal torque. The drag torque acts to close the gap due to the radial migration of dust particles from the outer disc (top centre panel).}
\label{fig:01mj1mm}
\end{center}
\end{figure*}
%
\section{Methodology}
\label{sec:methods}

\subsection{Dust and gas simulations}
We use the \textsc{phantom} SPH code \citep*{price10a,lodato10a,price12a,nixon13a} to perform global 3D simulations of dusty discs containing one or more embedded planets. The three processes involved in the gap-opening process --- gas viscosity, gravity and drag --- are computed consistently. Artificial viscosity is applied, calibrated to mimic the viscous transport of gas described with a Prandtl-like model of turbulence \citep{lodato10a}. The central star and the planets are modelled using sink particles. Sink particles are allowed to both accrete and migrate as a consequence of their interactions with the disc \citep*{bate95a} and their gravitational interaction. In particular the gravity between the sink particles and the gas and dust particles in the disc is modelled.



Drag between gas and dust is computed using the two-fluid algorithm described in \citet{laibe12a}. 
The algorithm has been extensively benchmarked on test problems with known analytic solutions \citep{laibe11a,laibe12a,price15a}. 
We additionally perform simulations where gravity from the planets is artificially turned off in the dust. This allows us to contrast our results with a model where the dust is assumed to respond to the gas only via the drag torque.

\subsection{Initial conditions}
We setup a disc as in \citet{lodato10a}. We assume a central star of mass $1.3\, M_{\odot}$ surrounded by a gas disc made of $5\times10^{5}$ gas particles and a dust disc made of $3\times10^{5}$ dust particles. The two discs extend from $r_{\rm{in}}  = 1$ au to $r_{\rm{out}}= 120 $ au.  We model the initial surface density profiles of the discs using power-laws of the form $\Sigma(r) = \Sigma_{\rm in} (r/r_{\rm in})^{-p}$. We adopt $p= 0.1$ and set $\Sigma_{\rm in}$ such that the total gas mass contained between $r_{\rm in}$ and $r_{\rm out}$ is 0.0002 M$_{\odot}$. 
We assume 1 mm dust grains with a corresponding Stokes number (the ratio between the stopping time and the orbital timescale), ${\rm St}\sim 10$.
The initial dust-to-gas ratio is $0.01$ and ${\rm St} \propto 1/\Sigma_{\rm g} \sim r^{0.1}$ in the disc. We simulate only the inner part of the disc since this is what can be observed with ALMA e.g. in HL Tau. If the gas phase were to extend to $r_{\rm{out}}= 1000 $ au, the total mass of the system is $\simeq 0.01 $ M$_{\odot}$.  We assume a vertically isothermal equation of state $P=c_{\rm s}^{2} \rho$ with $c_{\rm s}(r) = c_{{\rm s,in}} (r/r_{\rm in})^{-0.35}$ and an aspect ratio of the disc that is $0.05$ at 1 au. We set an SPH viscosity parameter $\alpha_{\rm AV} = 0.1$ giving an effective \citet{shakura73a} viscosity $\alpha_{\rm SS} \approx 0.004$. We setup a planet located at 40 au and evolve the simulations over 40 planetary orbits. This is sufficient to study the physics of dust gap opening with our assumed grain size, though we caution that further evolution occurs over longer timescales. We vary the planet mass in the range $[0.05, 0.1, 0.5, 1] \,M_{\rm J}$ to evaluate the relative contributions of the tidal and drag torques. 
%

\section{Results}
\label{sec:results}
 Gap formation is a competition between torques. In a gas disc the competition is between the tidal torque from the planet trying to open a gap and the viscous torque trying to close it. Dust, by contrast, is pressureless and inviscid, and the competition is between the tidal torque and the aerodynamic drag torque.
 
Dust efficiently settles to the midplane in our simulations, forming a stable dust layer with dust to gas scale height ratio of $\sim\sqrt{\alpha_{\rm SS}/\mathrm{St}}\sim 0.02$, consistent with the \cite{dubrulle95a} model and other SPH simulations of dusty discs \citep[e.g.][]{laibe08a}. Settling of grains is expected to slightly reinforce the contribution from the tidal torque by local geometric effects.
  

\subsection{Mechanism I --- low mass planets}
Fig.~\ref{fig:01mj1mm} demonstrates gap-opening when the planet is not massive enough to carve a gap in the gas disc. The gas shows only a weak one-armed spiral density wake supported by pressure, as predicted by linear density wave theory \citep{ogilvie02a}.

The general expression for the drag torque is
\begin{equation}
\Lambda_{\rm d} = - r \frac{K}{\rho_{\rm d}} (v_{\rm d}^{\phi} - v_{\rm g}^{\phi}),
\end{equation}
 where $K$ is the drag coefficient, $\rho_{\rm d}$ is the dust density and $v_{\rm d}^{\phi}$ and $v_{\rm g}^{\phi}$ are the azimuthal velocity of the dust and gas, respectively.   For large grains and in stationary regime, the dust and the gas velocities are given by eq. 2.12-2.14 of \citet{nakagawa86a} (assuming corrections due to gas viscosity are negligible). Hence $\Lambda_{\rm d}\propto (\mathrm{St} + \mathrm{St}^{-1})^{-1}$ since it is dominated by the contribution from the background radial pressure gradient \citep{nakagawa86a}. 
In a disc undisturbed by a planet the radial pressure gradient and the gas viscous velocity are, in general, negative. Hence the drag torque on the dust phase is negative, pushing dust particles inwards. The planet exerts a tidal torque on the dust which is positive outside the planet orbit and negative inside, forcing grains away from the planet. This torque increases as the planet mass increases. The balance between these two torques determines the evolution of the dust: A gap opening process resisted by drag from the outer disc. 

At radii smaller than the orbital radius of the planet, both contributions add to make the grains drift inwards. At larger radii the tidal torque resists the drag torque by repelling the particles from the planet.  For any planet mass, the tidal torque therefore decelerates dust particles outside the planetary orbit. A strong enough tidal torque in the outer disc prevents the replenishment of the dust population in the inner disc due to the action of the drag torque. Gap closing (opening) induced by drag outside (inside) the planetary orbit is maximal for grains with $\mathrm{St} = 1$, decreases linearly with Stokes number for $\mathrm{St}\ll1$ and is inversely proportional to Stokes number for $\mathrm{St}\gg1$ \citep{takeuchi01a}. However, small grains follow the viscous evolution of the gas. We therefore expect a less effective replenishment of larger particles from outside the planetary orbit due to the reduced drag torque, resulting in a more symmetric gap around the planet location.


 Whether or not a gap can be opened in the dust depends on the planet mass. For a planet of very low mass (top row in Fig.~\ref{fig:01mj1mm}; 0.05 $M_{\rm J}$), we only see a partial depletion of dust at the planet location, while the regions of the disc inside the planetary orbit are constantly replenished by an inflow of particles from the outer disc. Replenishment occurs because the drag torque produces a constant inward drift of dust particles, while the action of the the tidal torque is localised at the planet position. If the time taken by dust particles to refill the gap is shorter than the orbital timescale, the gap (and inner disc) is refilled by the action of the drag torque. However, if the tidal torque from the planet on the inner disc is strong enough to overcome this refilling, an inner cavity will slowly open.

\begin{figure}
\begin{center}
\includegraphics[width=0.42\textwidth]{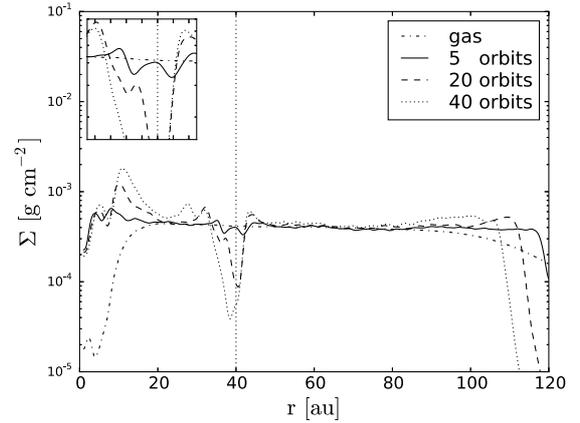}
\caption{Evolution of the azimuthally averaged dust surface density profile for the disc hosting a 0.1 $M_{\rm J}$ planet corresponding to the bottom centre panel of Fig. \ref{fig:01mj1mm}. The dotted-dashed line provides the gas surface density scaled by a factor of 0.01 at the end of the simulation, for direct comparison with the dust phase. The dotted vertical line indicates the planet orbit. An asymmetric gap profile is seen in the dust, but not in the gas. }
\label{fig:sigmatime}
\end{center}
\end{figure}

 The lower panel in Fig.~\ref{fig:01mj1mm} shows that a sufficiently massive planet (here 0.1 $M_{\rm J}$) clears a complete orbit in the dust. This occurs once the tidal torque is large enough to overcome the drag torque outside of the planetary orbit. 
 Fig.~\ref{fig:sigmatime} shows that the gap is not centred around the planet orbit, but is shifted towards the inner disc. Initially the gap is symmetric around the planet location, but becomes more and more asymmetric as the dust in the inner part of the disc drifts inwards, away from the planets orbit. As a result, no dust population is maintained in the corotation region. We confirm our physical picture above by noting that our 0.05 $M_{\rm J}$ simulation shows the same gap structure after a further 20 orbits, while the gap in the 0.1 $M_{\rm J}$ simulation grows more asymmetric with time.
 
\begin{figure*}
\begin{center}
\subfloat{\includegraphics[width=0.32\textwidth]{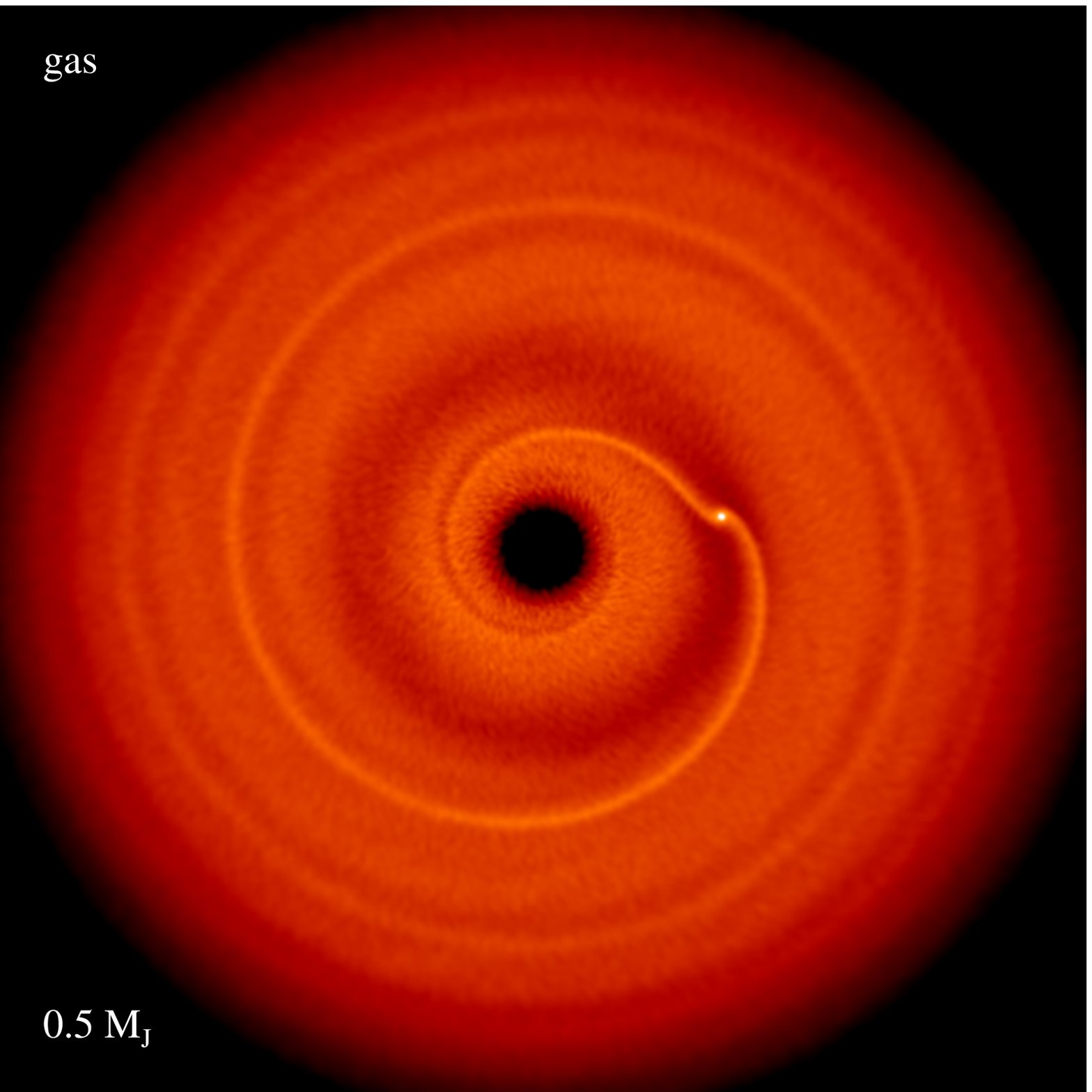}}
\hspace{0.025cm}
\subfloat{\includegraphics[width=0.32\textwidth]{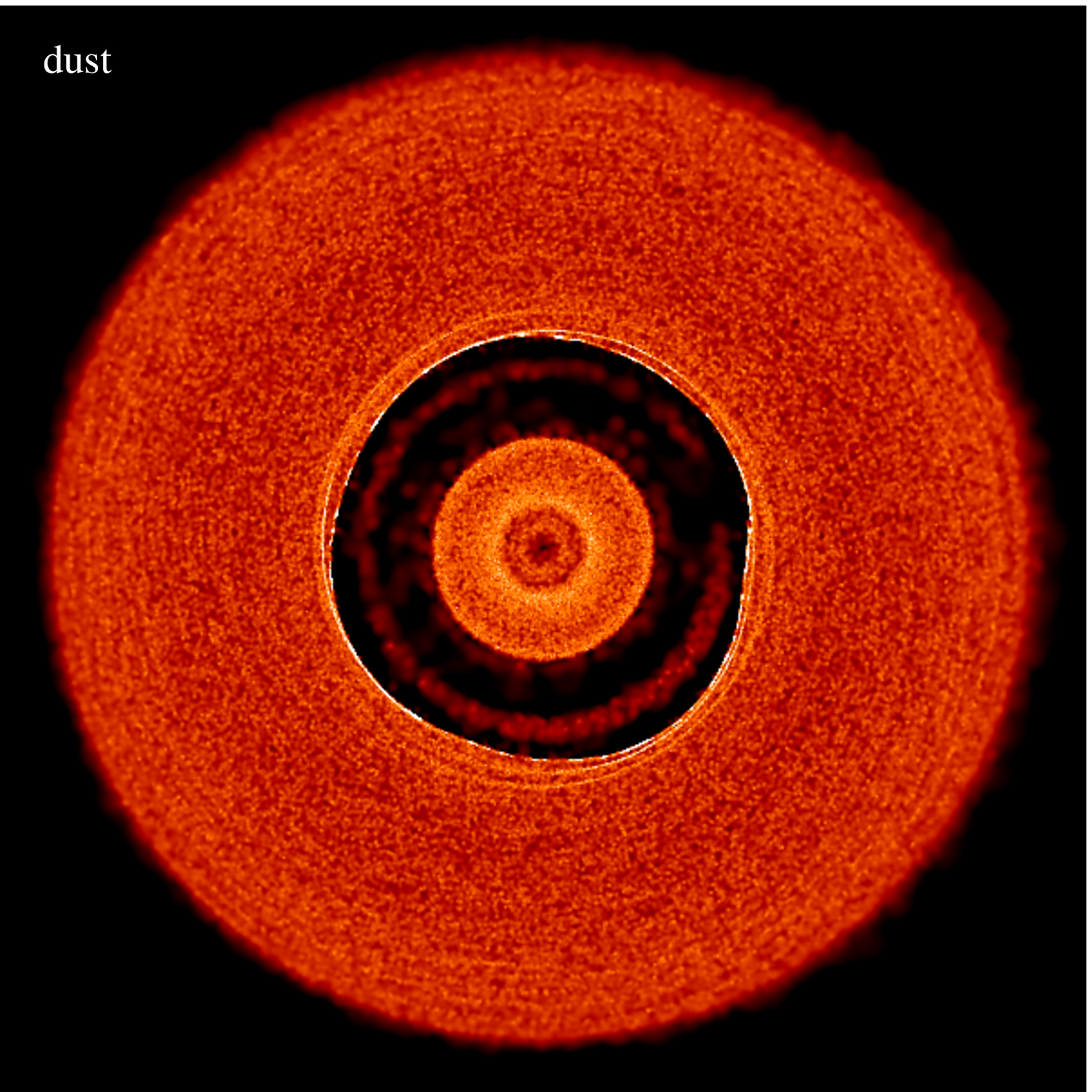}}
\hspace{0.25cm}
\subfloat{\includegraphics[width=0.32\textwidth]{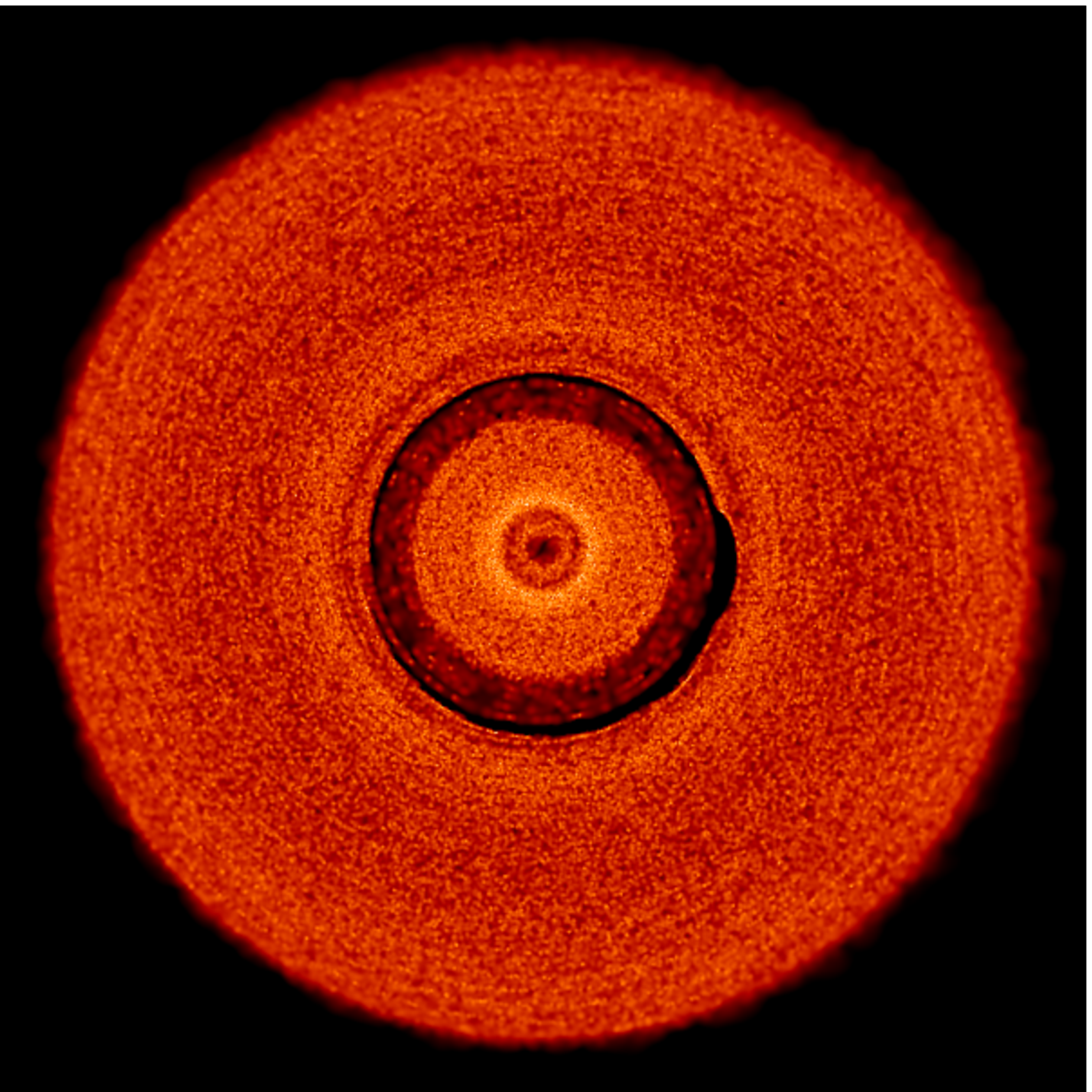}}
\vspace{-0.35cm}
\subfloat{\includegraphics[width=0.32\textwidth]{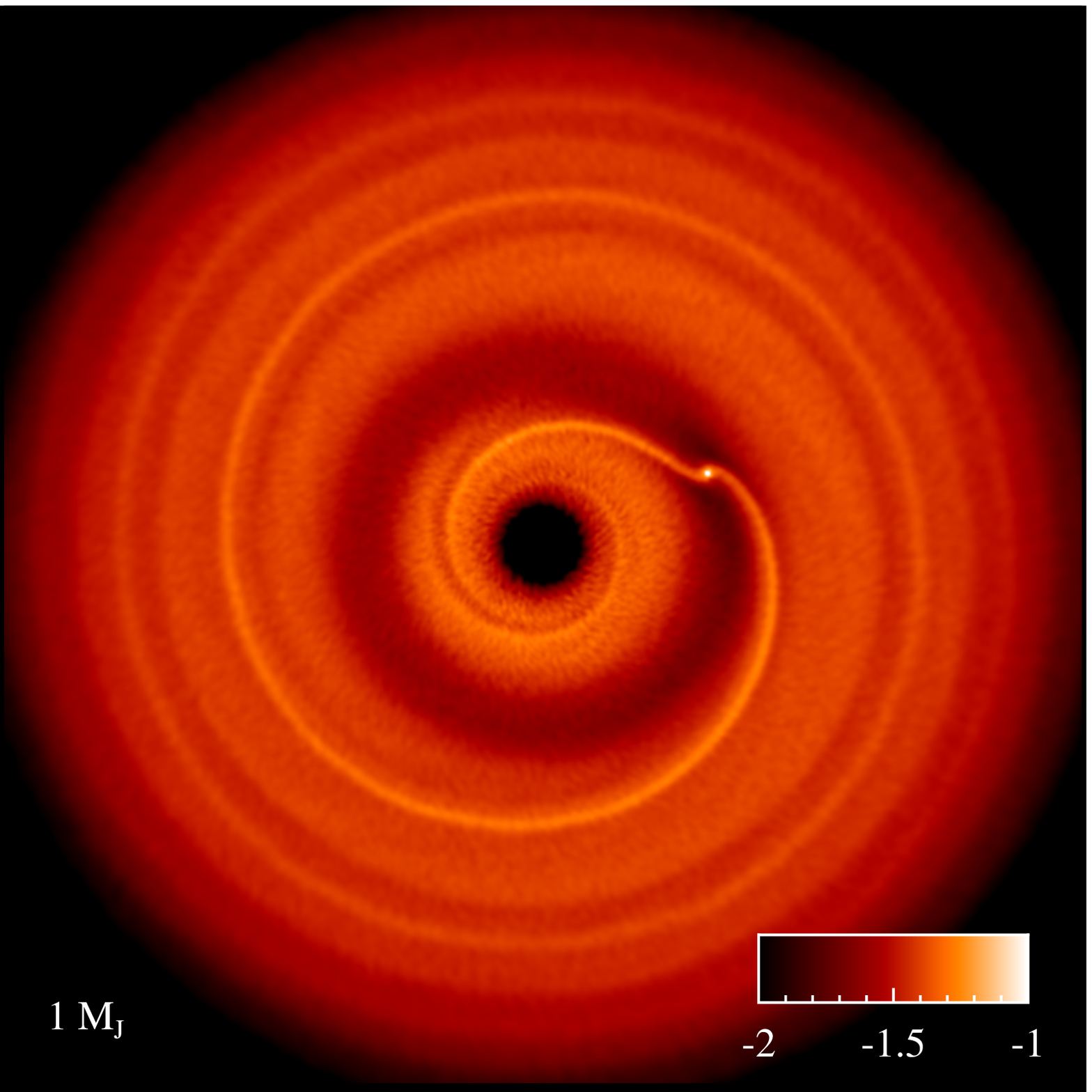}}
\hspace{0.025cm}
\subfloat{\includegraphics[width=0.32\textwidth]{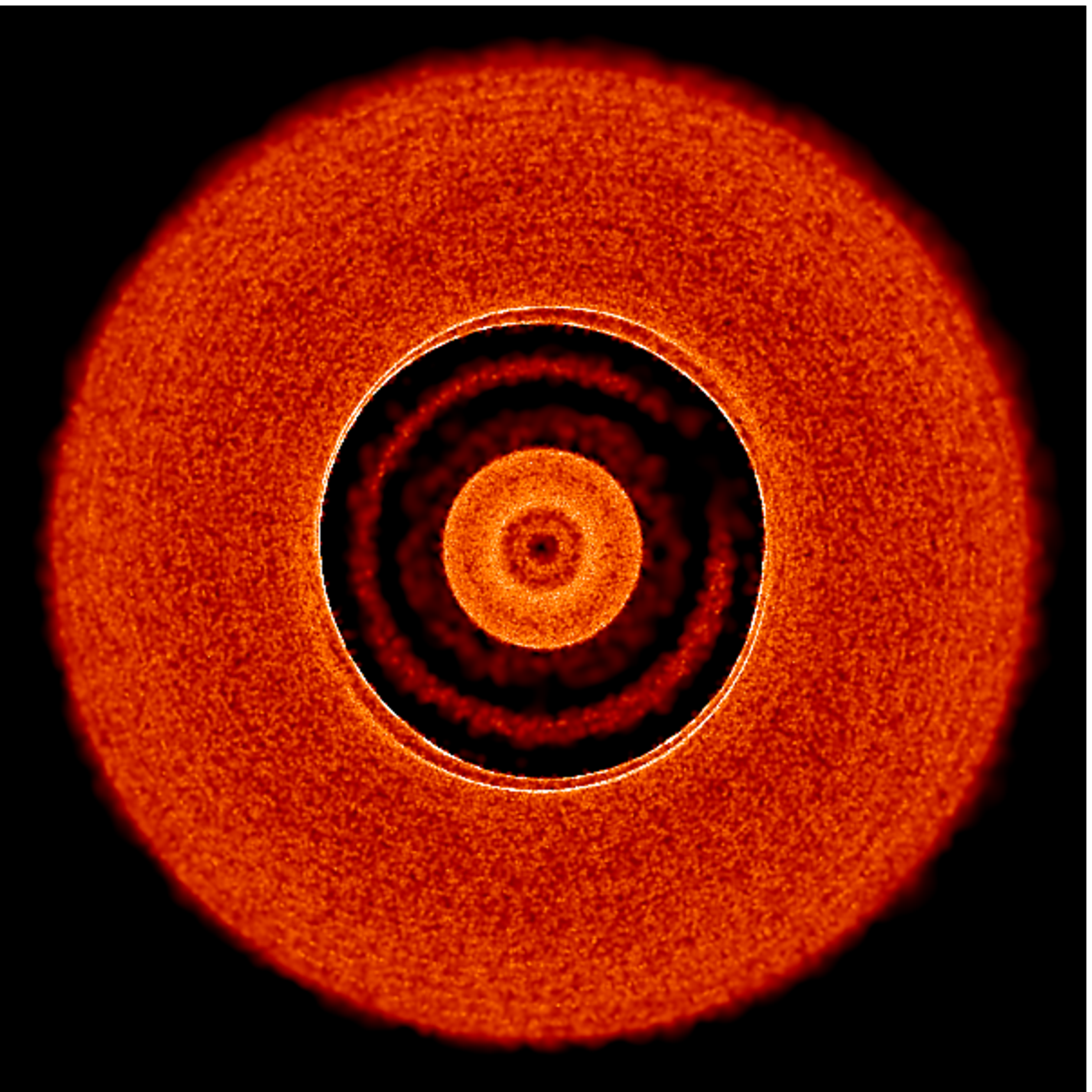}}
\hspace{0.25cm}
\subfloat{\includegraphics[width=0.32\textwidth]{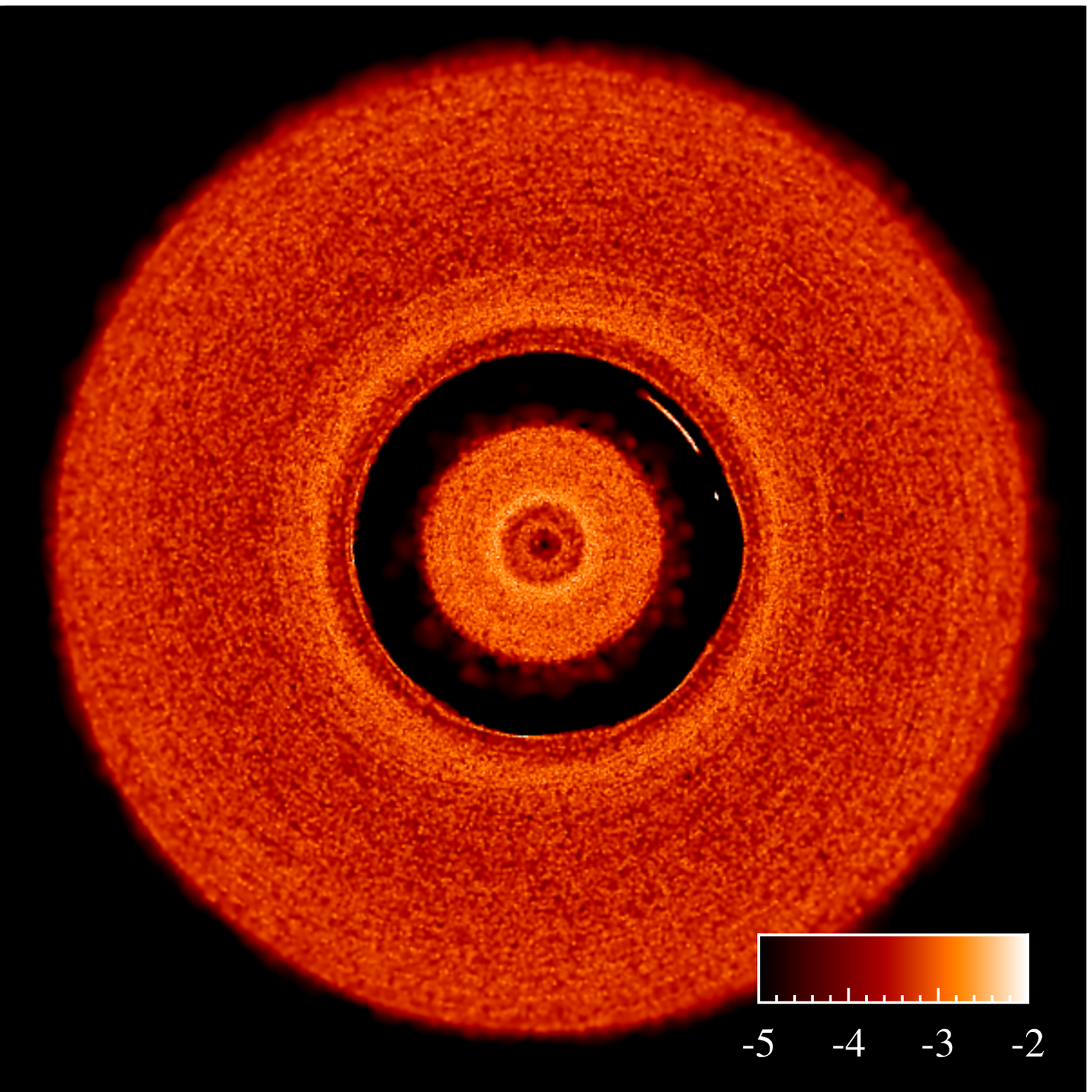}}
\caption{Gap opening via Mechanism II, where high mass planets carve a partial or total gap in the gas, and the dust is evacuated from the gap via drag and tidal torques. Plots are as in Fig.~\ref{fig:01mj1mm}, but with planet masses 0.5 $M_{\rm J}$ (top) and 1 $M_{\rm J}$ (bottom). Although the tidal torque modifies the structure of the gap and stabilises the corotation region, the structure in the dust phase is dominated by the drag torque (comparing centre and right panels).}
\label{fig:1mj1mm}
\end{center}
\end{figure*}

The right panel of Fig.~\ref{fig:01mj1mm} demonstrates the role played by the tidal torque in the gap-opening mechanism, showing the dust dynamics computed with the gravitational force from the planet on the dust phase switched off. In this case, solid particles are unaffected by the surrounding gas and simply migrate towards the central star.

%

\subsection{Mechanism II --- high mass planets} 

\begin{figure*}
\begin{center}
\includegraphics[width=0.34\textwidth,]{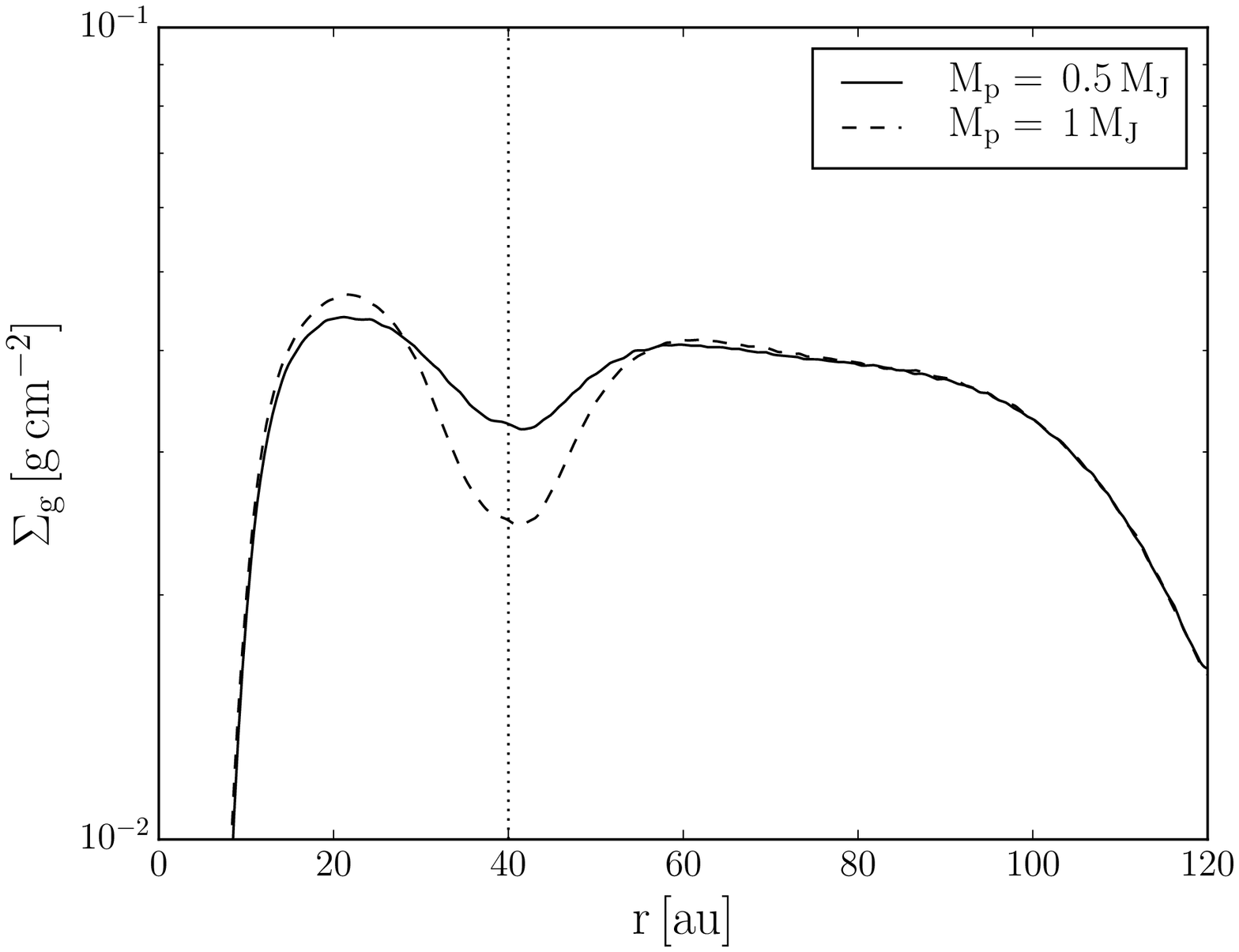}
\includegraphics[width=0.34\textwidth]{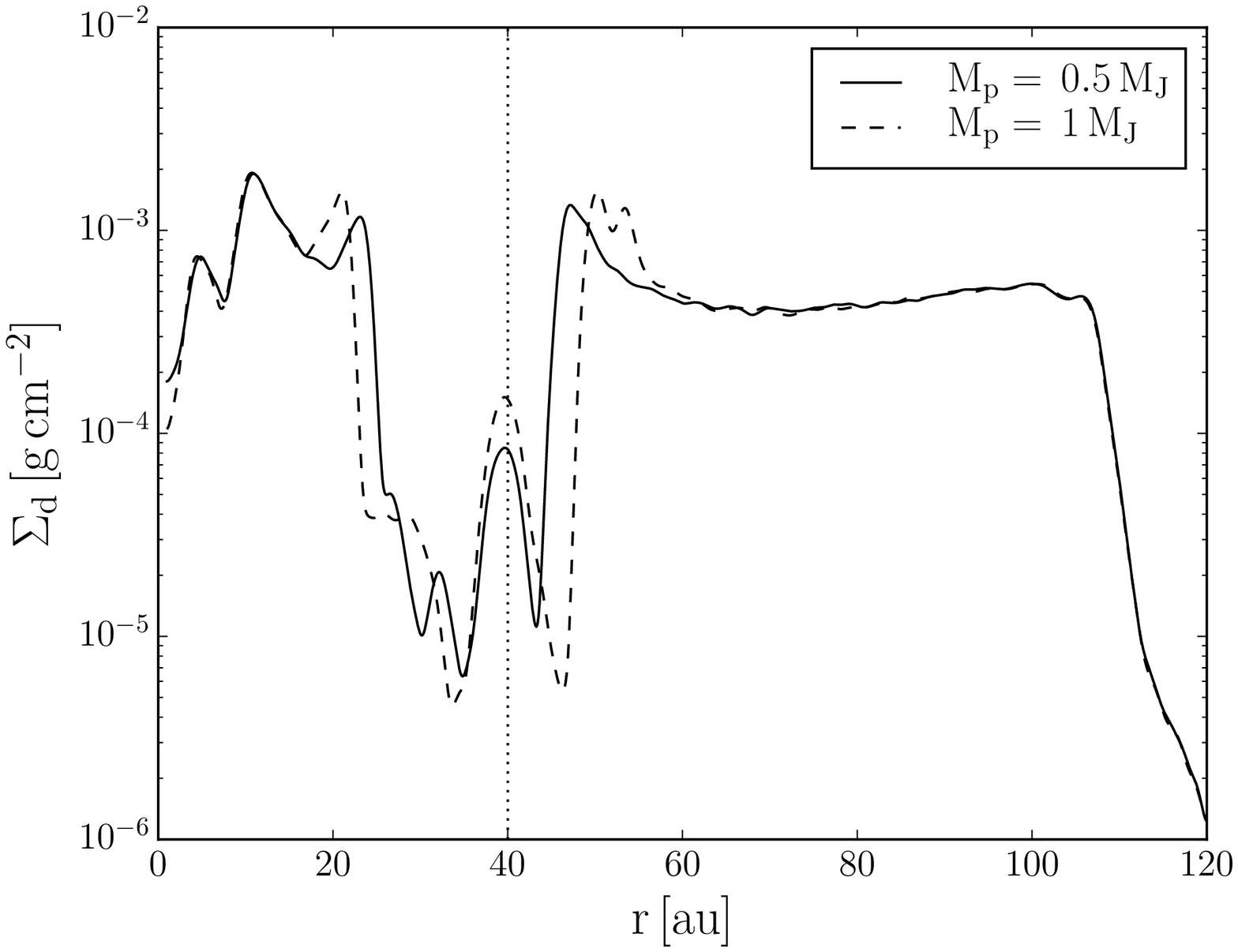}
\includegraphics[height=0.258\textwidth]{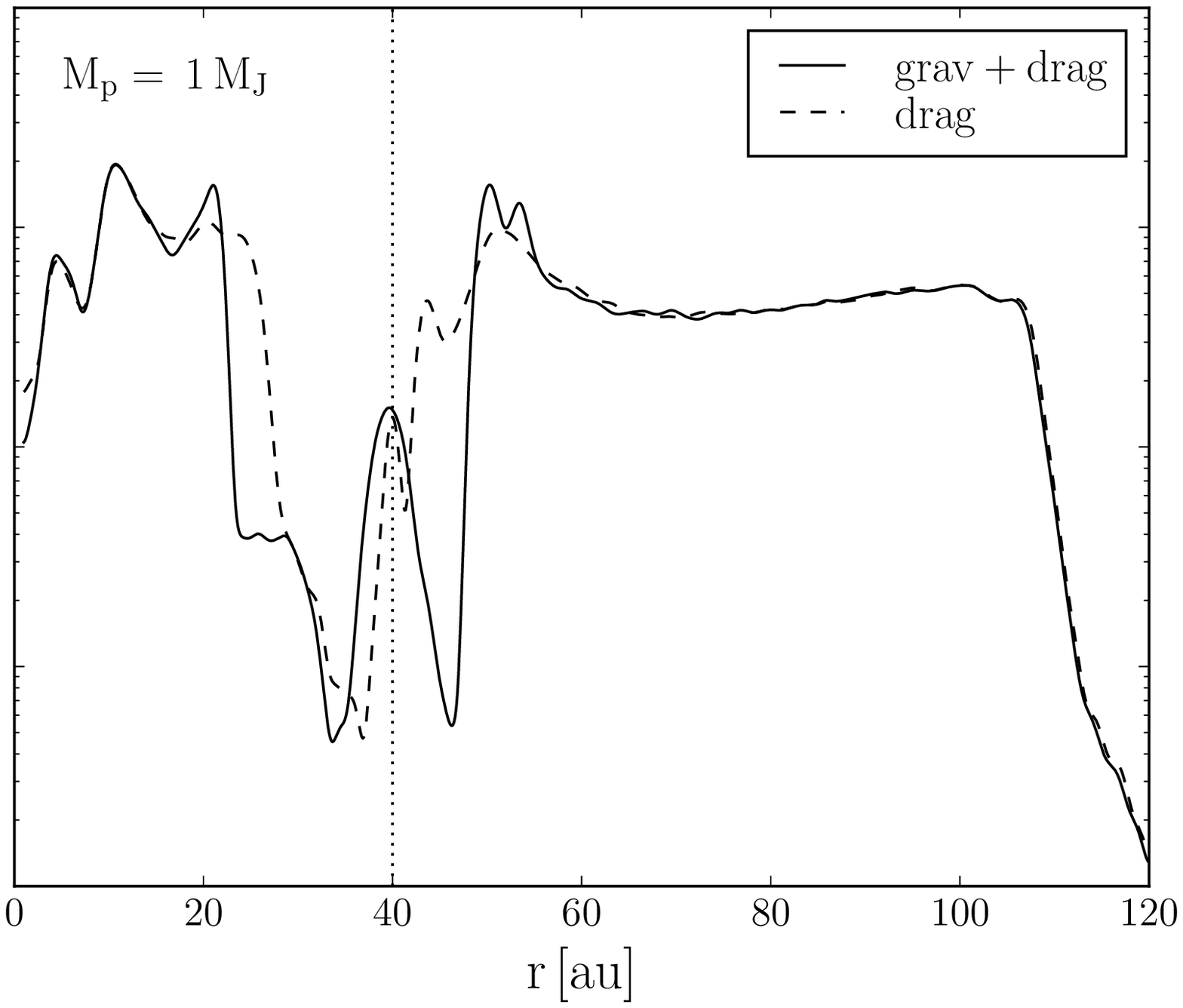}
\caption{Azimuthally averaged surface density of the gas corresponding to the simulations of Fig~\ref{fig:1mj1mm}. The dotted vertical line indicates the planet's location. Gaps are created in both the gas and the dust phases.}
\label{fig:sigmaall}
\end{center}
\end{figure*}
\begin{figure*}
\begin{center}
\subfloat{\includegraphics[width=0.32\textwidth]{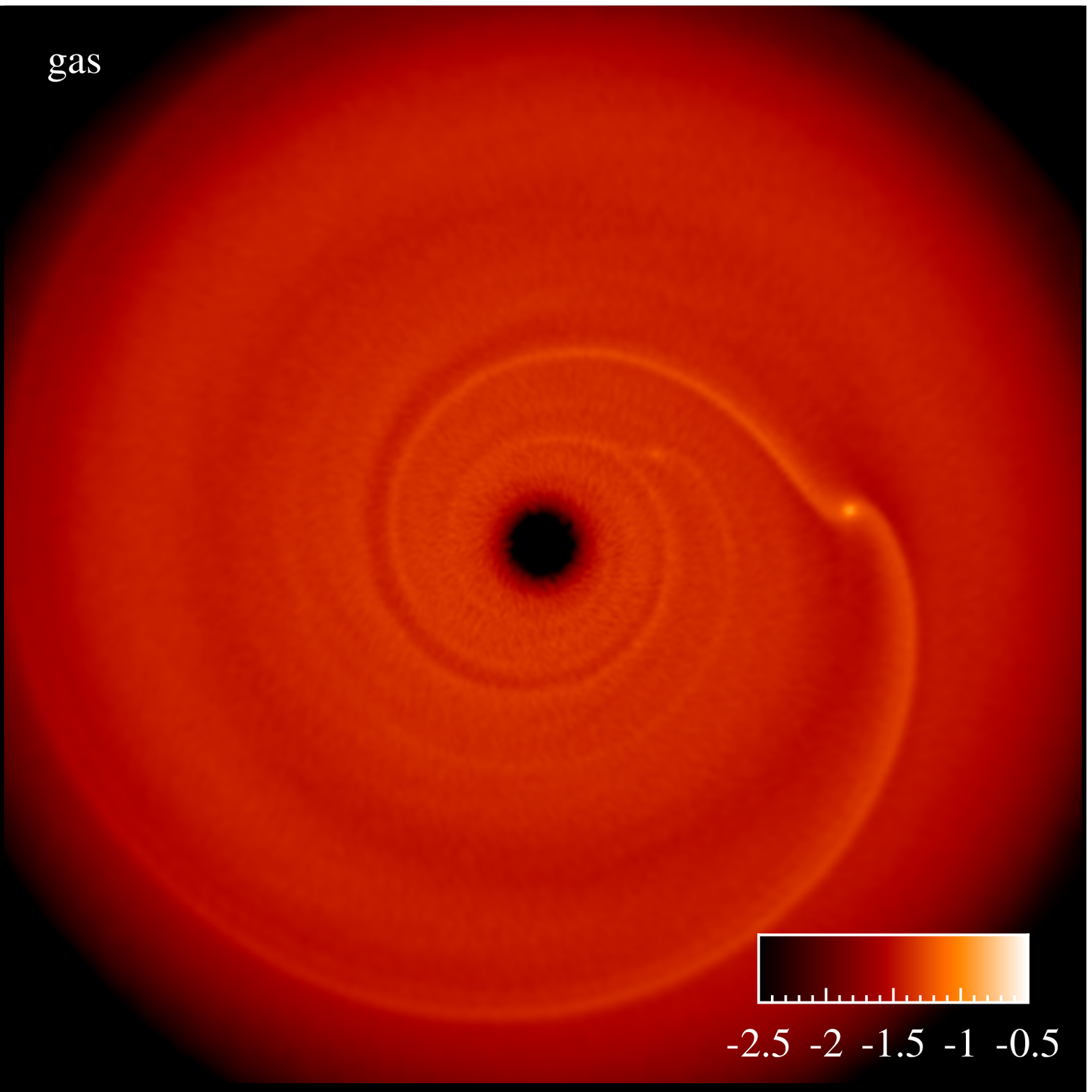}}
\hspace{0.025cm}
\subfloat{\includegraphics[width=0.32\textwidth]{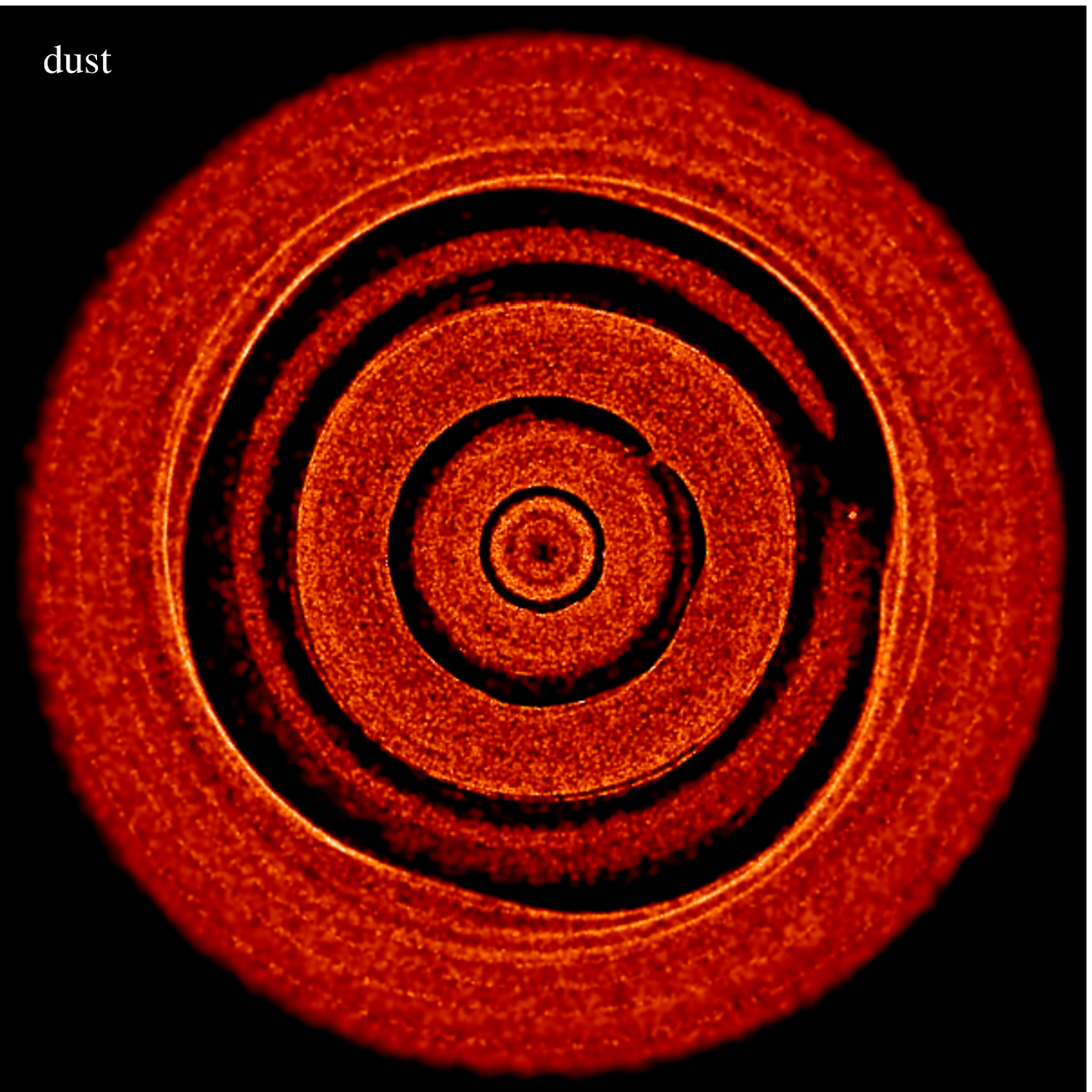}}
\hspace{0.25cm}
\subfloat{\includegraphics[width=0.32\textwidth]{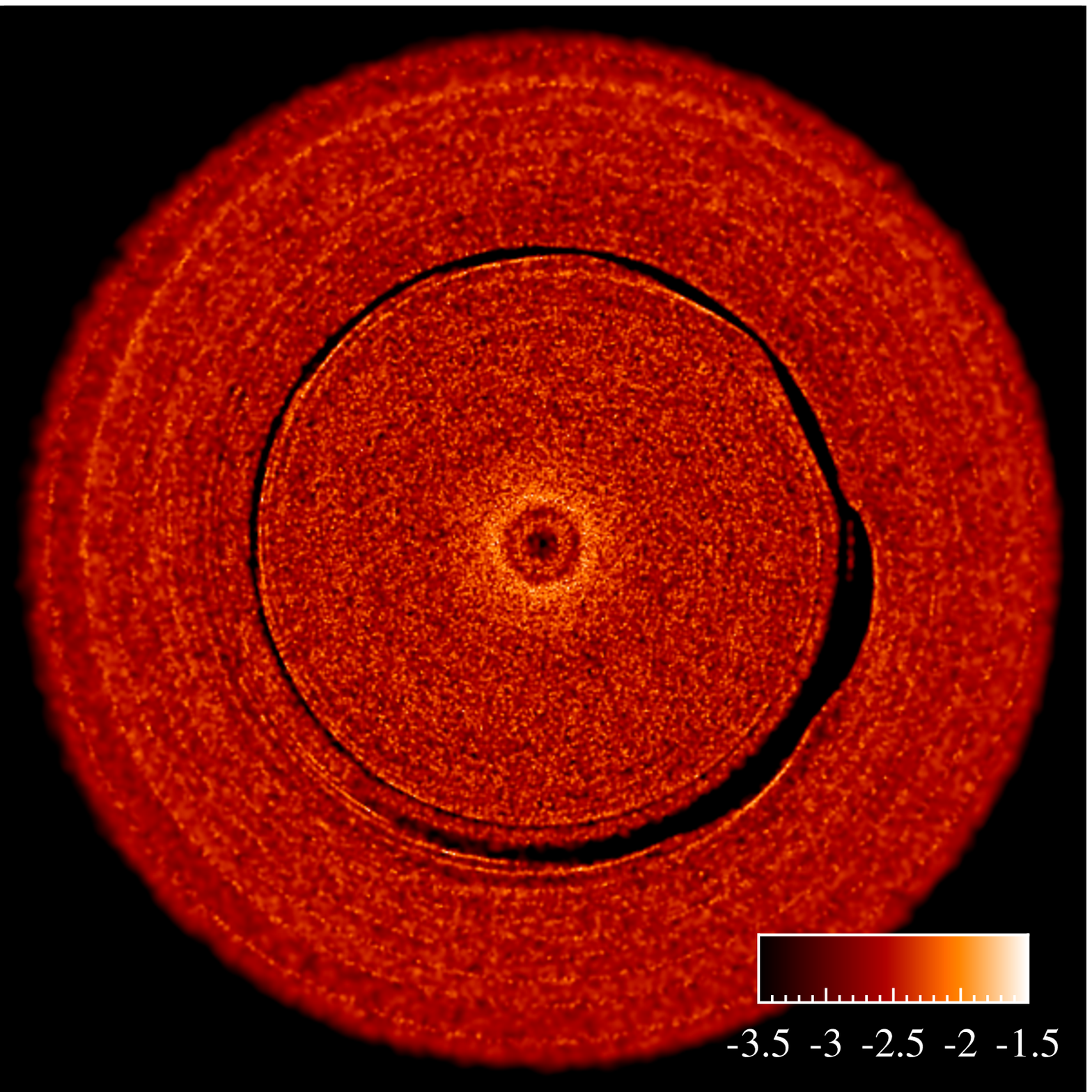}}
\caption{Similar to Fig.~\ref{fig:01mj1mm}, but with a disc hosting three planets of mass 0.08, 0.1 and 0.52 $M_{\rm J}$ initially located at the same distance as the gaps detected in HL Tau. The first two planets open dust gaps according to Mechanism I, whereas the third planet is at the transition between the two mechanisms. Importantly, the tidal torque in the dust can not be neglected.}
\label{fig:hltau}
\end{center}
\end{figure*}
Previous authors have identified two regimes for dust gap formation: high-mass planets opening gas and dust gaps (e.g. \citealt{crida06a}) and low-mass planets creating mild radial pressure gradients in the gas but deeper gaps in the dust (e.g. \citealt{paardekooper04a}). Both of these regimes involve the same, drag assisted, gap opening mechanism (our Mechanism II), but with different intensities. For sufficiently massive planets (0.5 $M_{\rm J}$ and 1 $M_{\rm J}$, top and bottom rows, respectively), the tidal torque dominates the viscous torque at the planet location, expelling the gas. This results in a stable gap in the gas surface density. 
A pressure maximum is always created at the outer edge of the gap. At the inner edge, a pressure maximum forms only if the perturbation of the pressure profile induced by the tidal torque exceeds the background pressure gradient. This is more likely to occur for flat pressure profiles and a planet located close to the star \citep{crida07a,fouchet10a}. The drag torque on the dust phase is influenced by the formation of these pressure maxima at the gap edges. Dust accumulates at the location of the pressure maxima, forming a deeper dust gap than in the gas. 


 As expected, the width and depth of the gaps increase as the planet mass increases (top and bottom rows in Fig.~\ref{fig:1mj1mm}). Compared to the low mass planet case, the additional pile-up induced by the presence of pressure maxima concentrates dust more efficiently at the outer edges of the gap. Fig.~\ref{fig:sigmaall} shows that the gaps in the dust disc are W-shaped and asymmetric around the location of the planet. Dust grains located initially between the inner edge of the gap in the gas and the corotation radius are depleted towards the inner pressure maximum by the drag torque. In contrast to Mechanism~I, particles with $\mathrm{St}=1$ concentrate fastest on the gap edges. It is therefore easier for marginally coupled particles to open gaps in the dust \citep{weidenschilling77}.

A large and stable population of dust grains is observed in the corotation region where the tidal torque provides a stabilising effect. The efficiency of the drag torque decreases locally as the gas surface density decreases \citep{laibe12b} and the background pressure gradient is significantly reduced (see left panel of Fig.~\ref{fig:sigmaall}). Interestingly, excitation of particles eccentricities at the outer edge of the gap forms narrow ridges just outside the orbit of the planet \citep{ayliffe12a,picogna15a}. This effect does not occur inside of the orbital radius of the planet because the drag torque is strong enough to efficiently damp any resonances that develop \citep{fouchet07a,ayliffe12a}, similar to what occurs in the whole disc in Mechanism~I. For our 1 $M_{\rm J}$ planet, the outer edge of the dust gap is close to the 3:2 resonance ($r\sim52$ au), inducing a double peaked outer edge in the dust density profile (see Fig.~\ref{fig:sigmaall}).

The right panel of Figs~\ref{fig:1mj1mm} and \ref{fig:sigmaall} shows that for high mass planets, the formation of a gap in the dust can be recovered simply by considering drag effects and neglecting the action of the gravitational potential of the planet. However, the detailed structure of the gap is still different when the tidal torque is included: the gap is wider, deeper, with a corotation region, sharper edges and more asymmetries due to external resonances.

\subsection{Application to the HL Tau disc}

Based on the previous discussion, we can now interpret our modelling of HL Tau \citep{dipierro15a} in the context of the two different mechanisms for dust gap opening. In this model, we used a dust-to-gas ratio of $\sim 0.06$ to reproduce a $n(s)\propto s^{-3.5}$ grain size distribution. We setup three planets located at 13.2, 32.3 and 68.8 au, with masses 0.08, 0.1 and 0.52 $M_{\rm J}$, and accretion radii 0.25, 0.25 and 0.75 au respectively. Fig.~\ref{fig:hltau} shows the gas and the dust surface densities after 10 orbits of the third planet.
The first two planets open gaps via Mechanism I, with gaps in dust but not gas. The third planet has a mass at the transition between the two mechanisms, chaning the gas density slightly and carving a local gap which is rapidly filled by the viscous inflow of gas. Hence, we can interpret the gaps observed by ALMA with these two mechanisms (although alternative scenarios may work as well).

\vspace{-0.25cm}
\section{Conclusion}

We have identified two physical mechanisms for dust gap opening by embedded planets in dusty protoplanetary discs. Our conclusions are:
\begin{itemize}
\item[i)] Gap formation in the gas is not a necessary condition for opening a gap in the dust. 
\item[ii)] For low mass planets that do not produce pressure maxima in the gas, drag resists (assists) the tidal torque outside (inside) the planetary orbit, forming an asymmetric gap around the planet orbit.

\item[iii)] For high mass planets that create pressure maxima in the gas, solid particles are prevented from accreting onto the planet. Here drag assists the tidal torque, leading to the formation of a deep dust gap with a stable population of grains at the corotation region. 
\end{itemize}


\vspace{-0.25cm}
\section*{Acknowledgments}
We thank the referee for insightful comments. GD thanks Monash for CPU time on NeCTaR.  We acknowledge an ARC Future Fellowship and Discovery Project. GD and G. Lodato acknowledge funding via PRIN MIUR prot. 2010LY5N2T. G. Laibe is funded by ERC FP7 grant ECOGAL. We used SPLASH \citep{price07a}.

\label{lastpage}

\bibliography{biblio}

\end{document}